\documentclass[]{icas2022}

\def\BibTeX{{\rm B\kern-.05em{\sc i\kern-.025em b}\kern-.08em
    T\kern-.1667em\lower.7ex\hbox{E}\kern-.125emX}}

\graphicspath{{ICAS-2022/figs/}}

\TitlePaper{A bi-level co-design approach for multicopters}
\AuthorPaper[1]{Jordan Mabboux}
\AuthorPaper[1]{Scott Delbecq}
\AuthorPaper[1]{Valérie Pommier-Budinger}
\AuthorPaper[1]{Jo\"el Bordeneuve-Guibe}
\AuthorPaper[2]{Ana Truc-Hermel}
\AuthorPaper[2]{Jean-Marie Kai}
\affil[1]{ISAE-SUPAERO, Université de Toulouse, France}
\affil[2]{Safran Tech, Energy \& Propulsion Department, France}

\abstractEnglish{Over the last few decades, multicopters have become significantly popular because of their various applications.
However, multicopters suffer from poor endurance performance as compared to other aircraft configurations with lifting surfaces.
In this paper, a co-design approach for the multicopter structure and propulsive system is presented with the objective of improving the energetic efficiency of the vehicle, while respecting dynamic performance requirements. For this purpose, the approach uses a bi-level formulation to optimize concurrently the design and the control. The methodology offers the possibility to activate or not a passive fault tolerant strategy in the controller synthesis to be tolerant to single rotor failure. In contrast to conventional design approaches that sequentially design each part, the co-design approach leads to a better optimal solution due to the consideration of the interactions and couplings between design and control.
With the proposed co-design approach, the results show the soundness of the proposed design methodology and a significant reduction in the consumed energy compared to a reference non-optimal design, while maintaining the same requirements in terms of dynamic performance
.}
\keywords{Co-design, Design optimization, Multicopter, 
Fault tolerant control}

\begin{document}

\body  

\section{Introduction}
\label{sec:intro}
Over the last few decades, multicopters have significantly become popular because of their various applications partly because of their great maneuverability and ability to execute complex movements. However, multicopters suffer from poor endurance performance as compared to other aircraft configurations with lifting surfaces \cite{zhang2021performance}. \newline
The preliminary design of efficient and reliable engineering systems motivates the development of methods allowing to take into account, very early in the design process, the performances and constraints of the different components that compose the system and the control system \cite{allison2014multidisciplinary, delbecq2017framework}.  \newline 
For aerospace applications, emphasis is placed on safety and thus reliability is sought to be integrated in the design optimization processes \cite{acar2005reliability, recalde2018reliability, nam2018multistage}. Moreover, total mass and energy consumption are often criteria sought to be minimized in the design optimization of aerospace vehicles \cite{brevault2020multi, palladino2021preliminary}. \newline 
Furthermore, for vehicles that have high dynamics, the control laws are sought to be integrated in the design optimization process \cite{brevault2020multi}. In the case of a multicopter, the subsystems such as the actuators (motors and rotors),  the energy storage system with batteries, the  structure (in particular the arms) are generally considered in the overall aircraft design process \cite{ budinger2020scaling, delbecq2020efficient, dai2021design}. \newline
Co-design approaches that include the control laws in the design optimization process are available in a limited number and generally address fixed-wing aircraft configurations \cite{van2019co} and not multicopters. The integration of multiple disciplines complicates the implementation and resolution of the design problem \cite{martins2021engineering}. It can also introduce multidisciplinary couplings/interactions that must be solved for the system to be consistent \cite{martins2021engineering}. \newline 

\begin{figure}[htp]
    \begin{subfigure}{0.246\textwidth}
    \centering
    \includegraphics[width=0.6\textwidth]{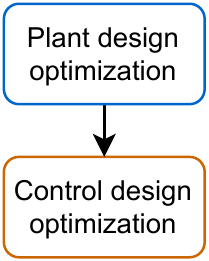}
    \caption{Sequential}
    \end{subfigure}
    \begin{subfigure}{0.246\textwidth}
    \centering
    \includegraphics[width=\textwidth]{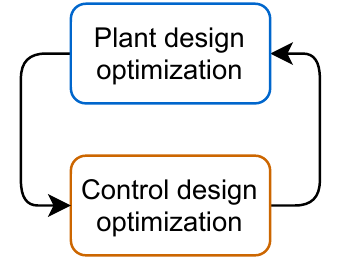}
    \caption{Iterative}
    \end{subfigure}
    \begin{subfigure}{0.246\textwidth}
    \centering
    \includegraphics[width=0.7\textwidth]{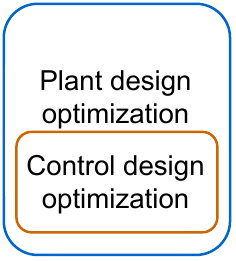}
    \caption{Bi-level (or Nested)}
    \end{subfigure}
    \begin{subfigure}{0.246\textwidth}
    \centering
    \includegraphics[width=0.72\textwidth]{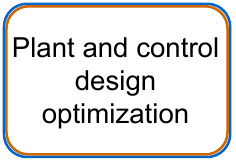}
    \caption{Simultaneous}
    \end{subfigure}
    \caption{Control and design optimization strategies}
    \label{fig:codesign_opti_strat}
\end{figure}

Co-design, also known as integrated plant and controller design, reduces the iterations of the traditional sequential approach because the constraints of each discipline are taken into account very early in the search for solutions, especially those related to control. Integrated plant and controller design strategies can be classified into four classes \cite{fathy2001coupling}, represented in figure~\ref{fig:codesign_opti_strat}.

\begin{itemize}
    \item \textbf{Sequential:} the plant design is optimized then the control design is optimized. The two distinct tasks are treated sequentially. This is the conventional design in engineering applications.
    \item \textbf{Iterative:} same procedure as previously, except that the process is iterated several times.
    \item \textbf{Bi-level (or Nested):} in this approach two nested optimization loops are used.
    \item \textbf{Simultaneous:} plant and control design are optimized simultaneously.
\end{itemize}

Sequential and iterative approaches lead to a sub-optimal solution if cross terms exist as stated in \cite{fathy2001coupling,reyer2002combined,bhattacharya2021control}. Simultaneous approach provides an optimal solution by integrating plant and control system design in the same loop. However it can be hard to solve this problem numerically. An alternative would be to formulate the problem equivalently with a bi-level nested approach that employs distributed optimization. \newline

In this paper, a bi-level co-design approach for multicopters is proposed. The control laws is tuned with a robust $H_\infty$ control strategy allowing the satisfaction of the dynamic performance requirements of the vehicle. The design optimization of the vehicle is achieved using lightweight sizing models that enable to vary the main design variables of the propellers, the electric motors, the structure and the battery whilst respecting technological and performance constraints. The control and the design optimization process are then assembled to form a co-design process using a bi-level strategy. The process is fully implemented in Matlab and the plant model (multicopter) is used by both the control law synthesis and the design optimization sub-processes. Furthermore, in the co-design process it is possible to activate or not the possibility to include additional requirements on the faulty model to synthesize a passive fault tolerant control. \newline

To present this methodology, the paper is organized as follows. The sizing models of the multicopter are given in section~\ref{sec:sizing_mdl}. The general approach of the control strategy used in the co-design approach, i.e. the structured $H_\infty$ synthesis where the tunable parameters are the position, attitude and velocity control gains for controlling the multicopter is briefly describes in section~\ref{sec:control}. The methodology and the general optimization problems are presented in section~\ref{sec:codesign_meth}. The first optimization results are presented in section~\ref{sec:codesign_res}. Finally, section~\ref{sec:concl} offers concluding remarks.

\newpage
\section{Sizing models}
\label{sec:sizing_mdl}

\begin{figure}[htp]
    \centering
    \includegraphics[width=7.5cm]{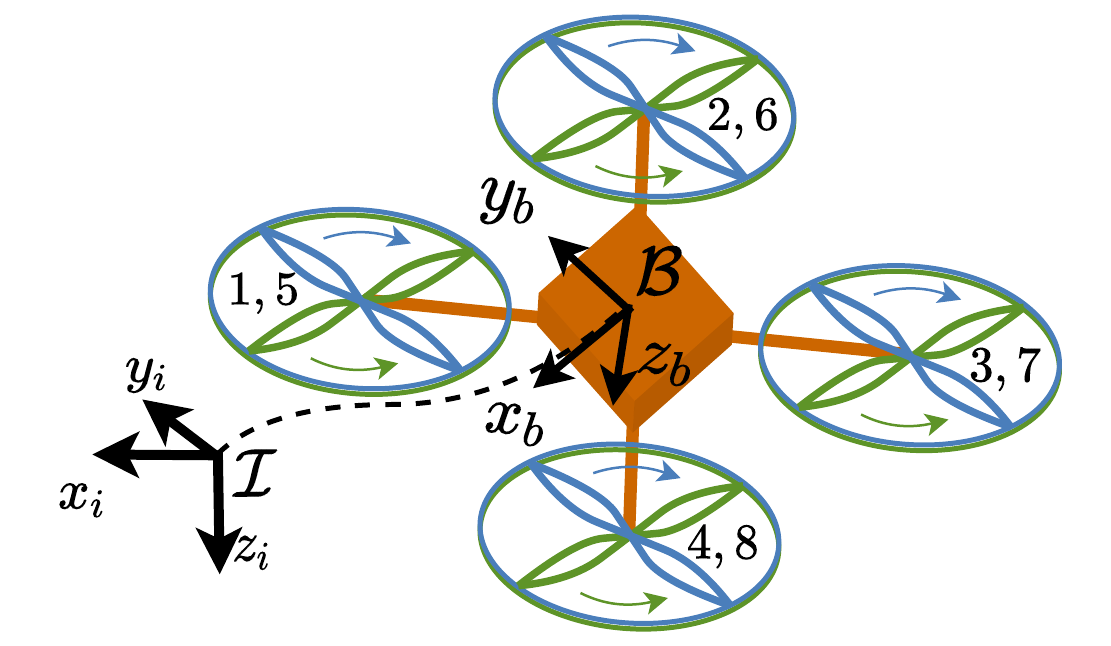}
    \caption{Quadcopter}
    \label{fig_Octocopter}
\end{figure}

The vehicle studied is an eVTOL quadcopter with two fixed pitch propellers per arm leading to a total of eight propellers. The structure is modeled as a central frame with composite alloy arms. The propulsion system is composed of propellers, electric motors, power converters, cables and a battery. An illustration is given in figure~\ref{fig_Octocopter}.

Sizing models are parametric models that predict the characteristics of a component or system according to design variables.
The main subsystems (motor, rotor, battery and arm) are designed through six design variables : the size of the motor, the rotor radius, the pitch rotor, the battery size, the arm length and the arm thickness, see figure~\ref{fig:archi_codesign_MC}. They are respectively stored in the vector of the plant design variables ($x^*_p$), see equation~\ref{eq_design_var}.

\begin{equation}
    x^*_p=[l^*_{mot}, r^*_{rot}, \theta^*_{0,rot}, v^*_{bat}, l^*_{arm}, e^*_{arm}]
    \label{eq_design_var}
\end{equation}

The stars $^*$ notation refers to the normalized value (scaling law factor). For example $l^*_{mot}=\frac{l_{mot}}{l_{mot,ref}}$ is the motor length of the new designed motor normalized by its reference. A part of these parameters is estimated thanks to scaling laws. The scaling laws of each subsystem are expressed in table~\ref{table_scale_laws}. The motor scaling laws come from~\cite{budinger2012estimation}, those of other subsystems are from physical expressions.

\begin{table}[htp]
    \begin{center}
        \begin{tabular}{c P{2.5cm} c c}
        \hline
        \textbf{Subsystems} & \textbf{Design variables $x_p$} & \textbf{Parameters} & \textbf{Estimation models} \\
        \hline
        \hline
        \multirow{ 7}{*}{\textbf{Motor}}
            &\multirow{ 7}{*}{$l^*_{mot}$} & Resistance & $R^*_{mot}=(l^{*}_{mot})^{-1} $\\
            & & Back EMF constant & $K^*_{e,mot}=(l^{*}_{mot})^{2}$ \\
            & & Viscous friction  & $K^*_{d,mot}=(l^{*}_{mot})^{3}$  \\
            & & Moment of inertia  & $J^*_{mot}=(l^{*}_{mot})^{5}$  \\
            & & Mass  & $m^*_{mot}=(l^{*}_{mot})^{3}$  \\
            & & Thermal resistance  & $R^*_{th,mot}=(l^{*}_{mot})^{-2}$ \\
            & & Thermal time constant  & $\tau_{th,mot}=\tau_{th,mot,ref}$ \\
            \hline 
            \multirow{5}{*}{\textbf{Rotor}}
            &\multirow{5}{*}{$r^*_{rot}$, $\theta^*_{0,rot}$} &Swept area & $S^{*}_{rot}=(r^{*}_{rot})^2$  \\
            & &Thrust coefficient & $c_{t0,rot}=f_{c_{t0,rot}}(\theta^{*}_{0,rot})$  \\
            & &Drag coefficient & $c_{q0,rot}=f_{c_{q0,rot}}(\theta^{*}_{0,rot})$  \\
            & &Moment of inertia & $J^{*}_{rot}=(r^{*}_{rot})^{5}$  \\
            & &Mass & $m^{*}_{rot}=(r^{*}_{rot})^{3}$ \\
            \hline 
            \multirow{ 3}{*}{\textbf{Battery}}
            &\multirow{3}{*}{$v^*_{bat}$} & Power & $P_{bat}=5 d_E.\frac{u_{cell,nom}}{u_{cell,max}} v^{*}_{bat}$  \\
            & & Mass & $m^{*}_{bat}= v^{*}_{bat}$ \\
            & & Energy & $E^{*}_{bat}=d_E.v^{*}_{bat}$ \\
            \hline 
            \multirow{ 3}{*}{\textbf{Arm}}
            &\multirow{3}{*}{$l^*_{arm}$, $e^*_{arm}$} & Diameter & $d^*_{arm}= f_{d,arm}(e_{arm})$\\
            & & Inertia & $J^*_{arm}= f_{J,arm}(e_{arm},l_{arm})$\\
            & & Mass & $m^*_{arm}= f_{m,arm}(e_{arm},l_{arm})$\\
        \end{tabular}
    \end{center}
    \caption{Scaling laws}
    \label{table_scale_laws}
\end{table}

\newpage
\section{Control structure}
\label{sec:control}
\begin{figure}[htp]
    \centering
    \includegraphics[width=0.65\columnwidth]{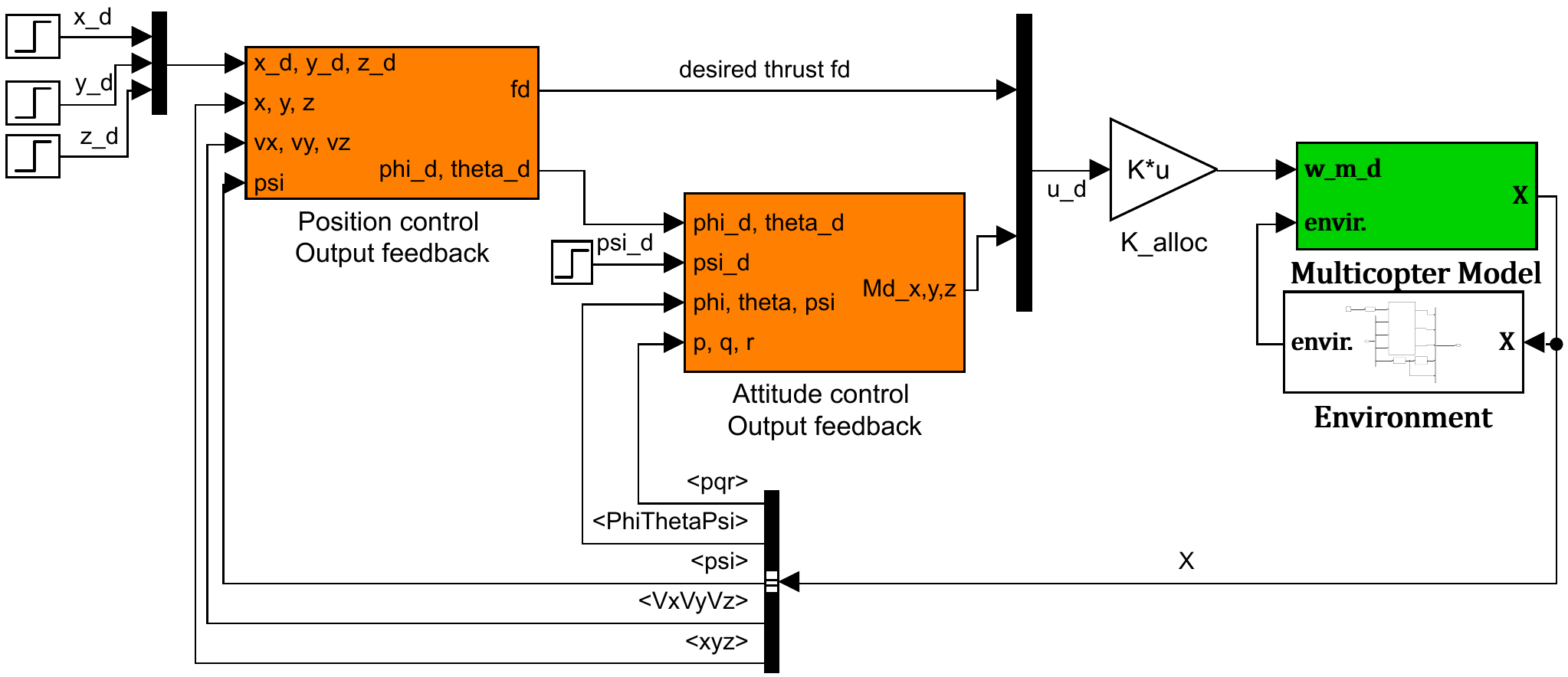}
    \caption{General structure of control for multicopter}
    \label{fig_ctrl_mdl_mc}
\end{figure}

The structure of the control can be seen in figure~\ref{fig_ctrl_mdl_mc}. Position and attitude control are both composed of three simple proportional controllers, one for each direction $x,y,z$ and $\phi,\theta,\psi$ respectively. The velocity and angular rate controllers are both composed of three simple proportional integral controllers. The controllers for the angular rates $p,q$  are composed of two simple proportional controllers. Therefore a total of 16 control gains must be tuned, see figure~\ref{fig:archi_codesign_MC}. 

Robust $H_\infty$ control is widely used to set the controller with a prescribed performance. Recent extensions of $H_\infty$ control allow the computation of fixed-structure control systems \cite{apkarian2012tuning}. In our application of co-design, this control strategy 
is selected since it is particularly well suited for computing the control gains by solving a non-smooth optimization problem subject to a variety of frequency-domain requirements. 

The $H_\infty$ fixed-structure control can also be used for the simultaneous co-design. The design variables and control gains are then optimized with the same optimization, as realized in \cite{denieul2018multicontrol}. Nonetheless, this optimization linearizes the model and cannot handle nonlinear constraints. Since the plant design constraints are nonlinear time-domain constraints which cannot be taken into account as such, plant design is performed through the outer optimization loop of the bi-level co-design strategy.

\begin{figure}[htp]
    \centering
    \includegraphics[width=0.45\columnwidth]{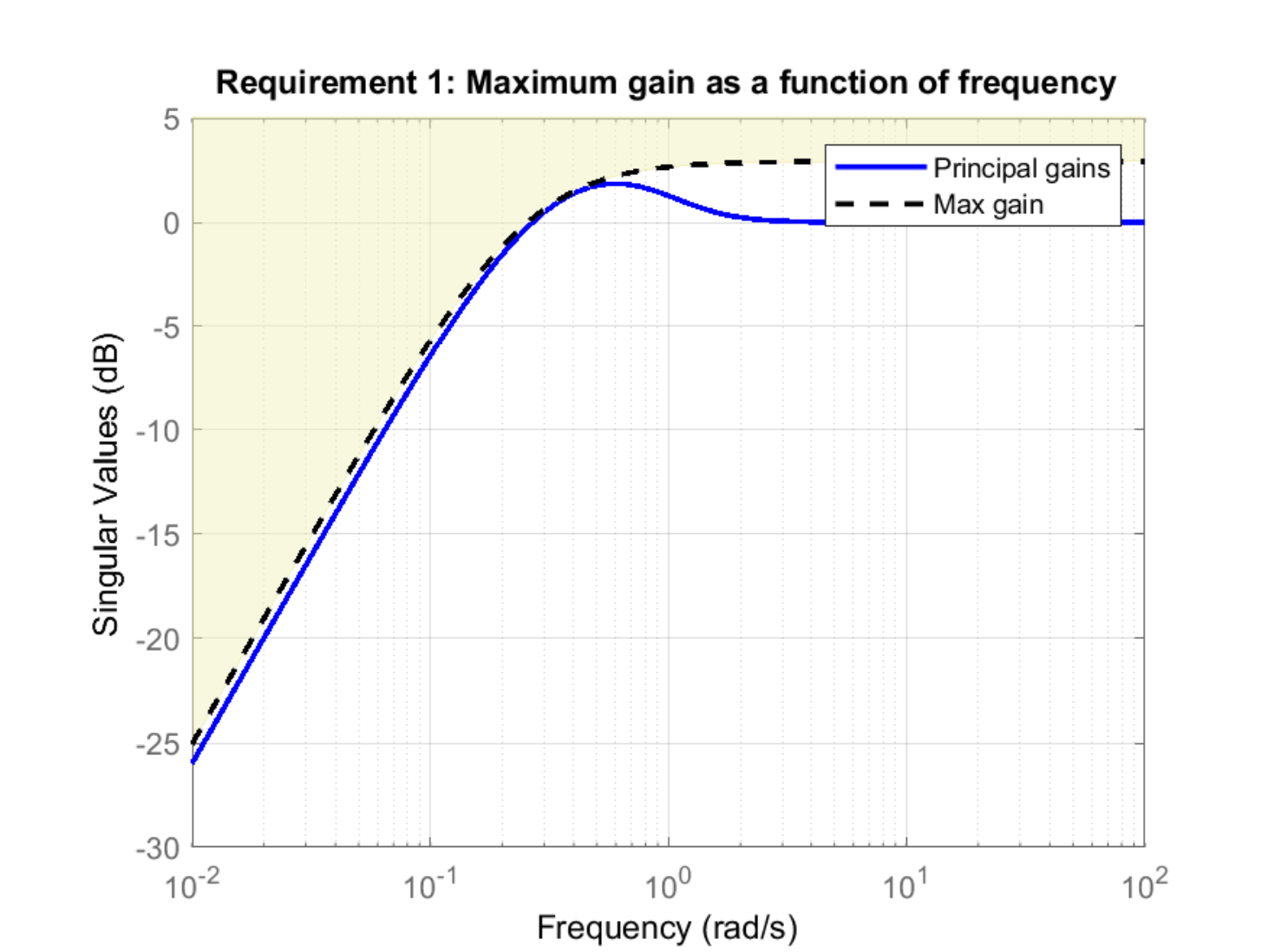}
    \caption{Output sensitivity function requirement}
    \label{fig:requirment_hinf_x}
\end{figure}

A strong advantage of 
the optimizer used is the ability to handle several models during the optimization process. This allows synthesizing a passive Fault Tolerant Control by integrating models with failure, e.g. single failure of any rotor as described in \cite{mabboux2021robust}. The model without failure must respect specified handling qualities requirements. In the same way, the set of all the models with a single rotor failure must respect specified requirements in degraded mode. Handling qualities requirements are specified using pole placement and constraints on the output sensitivity function $S(j \omega)$ defined by the transfer function between the reference and the error. This function must be small at low frequency to achieve good robustness and attenuation of disturbances. An example of requirements on the output sensitivity function is shown in figure~\ref{fig:requirment_hinf_x}.

\newpage
\section{Co-design methodology}
\label{sec:codesign_meth}
\begin{figure}[htp]
    \begin{subfigure}{0.5\textwidth}
    \centering
    \includegraphics[width=\textwidth]{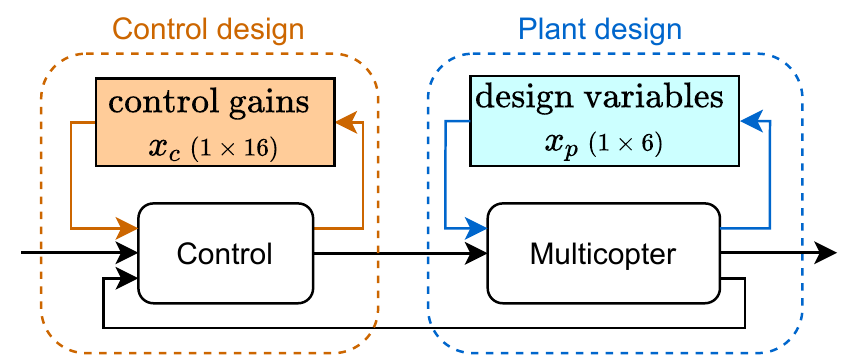}
    \caption{Co-design structure}
    \end{subfigure}
    \begin{subfigure}{0.49\textwidth}
    \centering
    \includegraphics[width=0.7\textwidth]{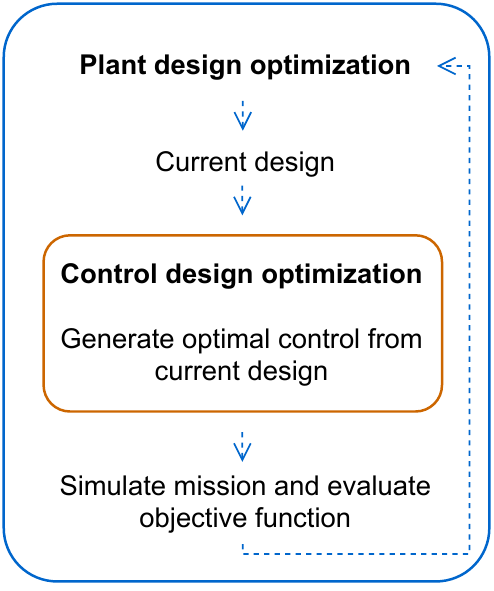}
    \caption{Bi-level optimization}
    \end{subfigure}
    \caption{Integrated plant and controller design: bi-level optimization}
    \label{fig:nested_opt}
\end{figure}

The structure of the control and plant design variables and the general workflow of the bi-level optimization are shown in figure~\ref{fig:nested_opt}. The bi-level co-design optimization includes two different optimization loops, used for their efficiency to solve the co-design problem. As stated in section \ref{sec:control}, the choice of using robust $H_\infty$ control leads to the need for a bi-level approach to include the vehicle design part. Indeed, simultaneous co-design strategy using only the $H_\infty$ optimization capability was not compatible with the integration of temporal response constraints as for example the maximum temperature for the thermal response of the electric motor. Furthermore, $H_\infty$ optimizer will be less suited to solve efficiently global optimization problem of plant design.

The two optimization loops are:
\begin{enumerate}
    \item The \textbf{system design optimization loop}: the outer optimization loop is described by problem~\ref{eq:outer_opt}.
    This optimization problem aims to minimize the energy consumed (or the total mass of the multicopter) with respect to the design variables defined in table~\ref{table_scale_laws} under motor, rotor, battery and arm constraints. 
    This optimization 
    optimizes the six plant design variables. This surrogate optimization is best suited for global minimization because the objective function takes a long time to evaluate since the nonlinear model of the multicopter must be fully simulated to evaluate the energy consumed, which is time-consuming. Seven design constraints are considered. The motor constraints include the maximum torque $T_{mot,max}$ and the maximum temperature $\Theta_{mot,max}$. The rotor limitations include the maximum rotational speed $\omega_{rot,max}$ and the maximum rotor radius. The first limitation is due to a limitation of the rotor tip speed. The second is a geometrical limitation to avoid overlap, which imposes a relation between rotor radius and arm length. The electrical source systems contains two different constraints. The last constraint is a structural constraint. The energy-mass loop is solved using the Normalized Variable Hybrid (NVH) formulation \cite{delbecq2020benchmarking}.
    
    \item The \textbf{control design optimization loop}: the inner optimization loop is defined by problem~\ref{eq:inner_opt}.
    The 
    optimizer is well suited to tune the 16 control gains with $H_\infty$ requirements on the linearized model in nominal case and models in degraded mode with rotor failures.
    The filter $W_1$ is chosen to reject output disturbances and the filter $W_2$ to attenuate the high frequencies of the control signal.
\end{enumerate}

\begin{equation}
    \begin{aligned}
        & \text{minimize}
        & & E_{mot} \text{ (or } m_{mc}) & 1\\
        & \text{with respect to}
        & &  x_p=[l^*_{mot}, r^*_{rot}, \theta^*_{0,rot}, v^*_{bat}, l^*_{arm}, e^*_{arm}] & 6 \\
        & \text{subject to} & &  \max\limits_{t,j} T_{mot,j}(t) -T_{mot,max} \leq 0 & 1\\
        & & & \max\limits_{t,j} \Theta_{mot,j}(t) - \Theta_{mot,max} \leq 0 & 1\\
        & & & \max\limits_{t,j} \omega_{rot,j}(t) - \omega_{rot,max} \leq 0 & 1\\
        & & & \text{rotor overlap constraint} & 1\\
        & & & \text{electrical source constraints} & 2\\
        & & & \text{structural constraint} & 1\\
        \end{aligned}
        \label{eq:outer_opt}
\end{equation}

\begin{equation}
    \begin{aligned}
        & \text{minimize}
        & & \NORM{T_{w\rightarrow z}(j\omega)}_\infty, \qquad \forall w\in\{x_d,y_d,z_d,\phi_d,\theta_d,\psi_d\} & 1\\
        & \text{with respect to} & & x_c=[k_{p,x}, k_{i,x}, ...] & 16  \\
        & \text{subject to}
        & & \NORM{W_1 S}_\infty  -1\leq 0 & 1\\
        & & & \NORM{W_2 K S}_\infty  -1\leq 0 & 1\\
        & & & \text{pole placement} & 6 \\
        \end{aligned}
        \label{eq:inner_opt}
\end{equation}

\begin{figure}[htp]
    \centering
    \includegraphics[width=\columnwidth]{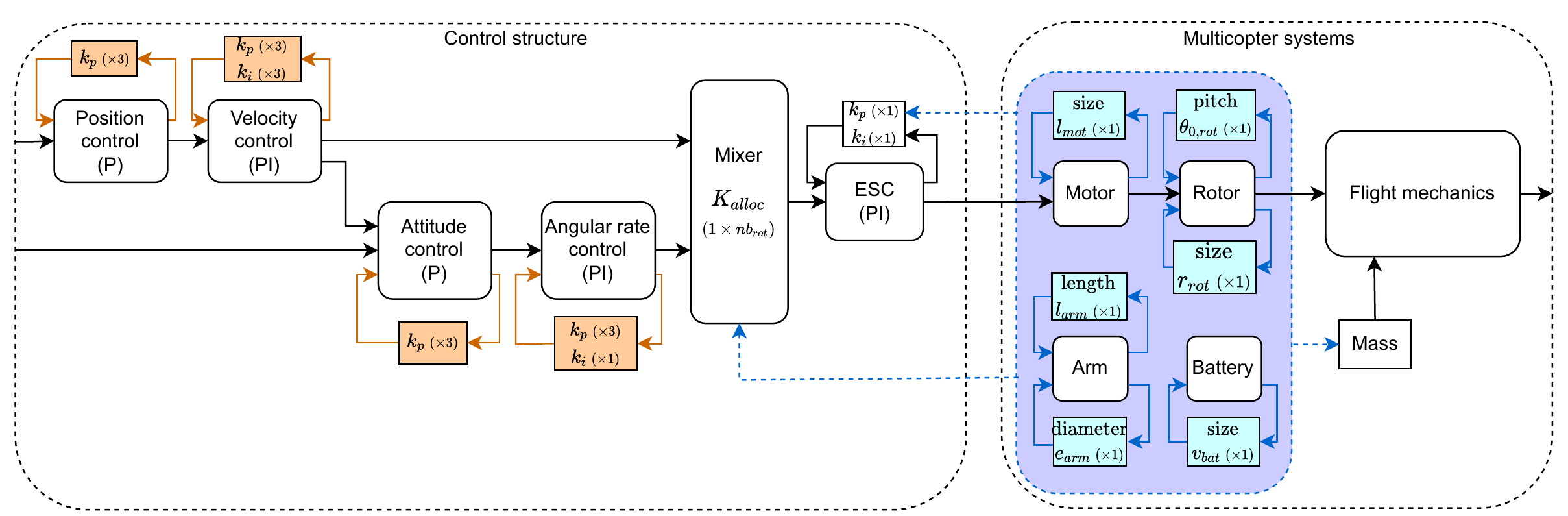}
    \caption{Detailed co-design structure}
    \label{fig:archi_codesign_MC}
\end{figure}

The co-design structure, detailed in figure \ref{fig:archi_codesign_MC}, is implemented and solved in Matlab.

The full computation time of the co-design process is around 20 hours on a typical scientific computing laptop \footnote{Intel Core i7-6820HQ CPU @ 2.70GHz, RAM 16Go} and can drop to 16 hours on a High Performance Computing cluster \footnote{2x Intel Xeon Gold 6126 CPU @ 2.66GHz, RAM 96Go} with no parallelization.

\section{Co-design results}
\label{sec:codesign_res}

\subsection{Energy and mass minimization}

Two co-design optimizations have been studied with  different objectives. The first one concerns a minimization of the consumed energy $E_{mot}$, while the second one is a reduction of mass $m_{mc}$. Both results are presented in table~\ref{tab:res_optim}. When the objective is to minimize the energy consumed, the latter is significantly reduced by 37\% compared to the reference, while the mass decreased by 10\%. When the objective is to minimize the mass of the multicopter, the optimized multicopter is lighter, the mass decreases by 14\%, but is less energy-efficient. The energy is only reduced by 14\%.

\begin{table}[htp]
    \begin{center}
        \begin{tabular}{c || c c c}
        &  $\min E_{mot}$ & $\min m_{mc}$ \\
        \hline
        \hline
        Energy consumed relative change [\%] & \textbf{-37} & -14\\
        \hline
        Total Mass relative change [\%] &
        -10 &
        \textbf{-14} \\
        \end{tabular}
    \end{center}
    \caption{Co-design optimization results}
    \label{tab:res_optim}
\end{table}

\begin{figure}[htp]
    \centering
    \includegraphics[width=0.75\textwidth]{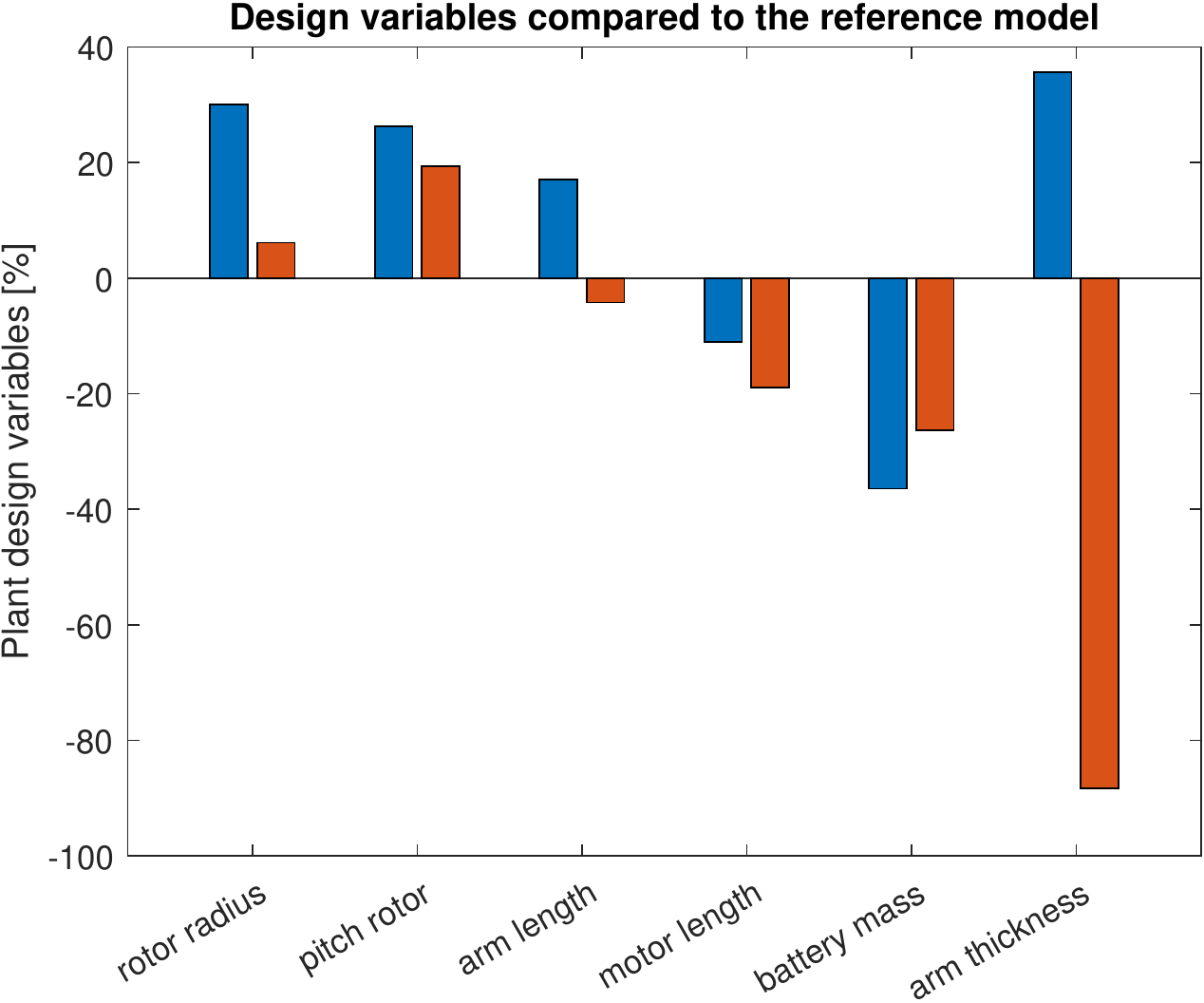}
    \caption{Design variables after optimization}
    \label{fig:design_opt}
\end{figure}

The plant design variables in relative change are plotted in figure~\ref{fig:design_opt} for both objective functions. With the exception of battery mass, the design parameters when mass is optimized, are smaller than those found when energy is minimized. This seems logical since the optimizer tries to reduce the size of each component as much as possible, when the objective is to decrease the total mass. However, this solution consumes more energy. Therefore it requires more battery, so the reduction in battery size is less important.

\begin{figure}[htp]
    \centering
    \includegraphics[width=0.85\textwidth]{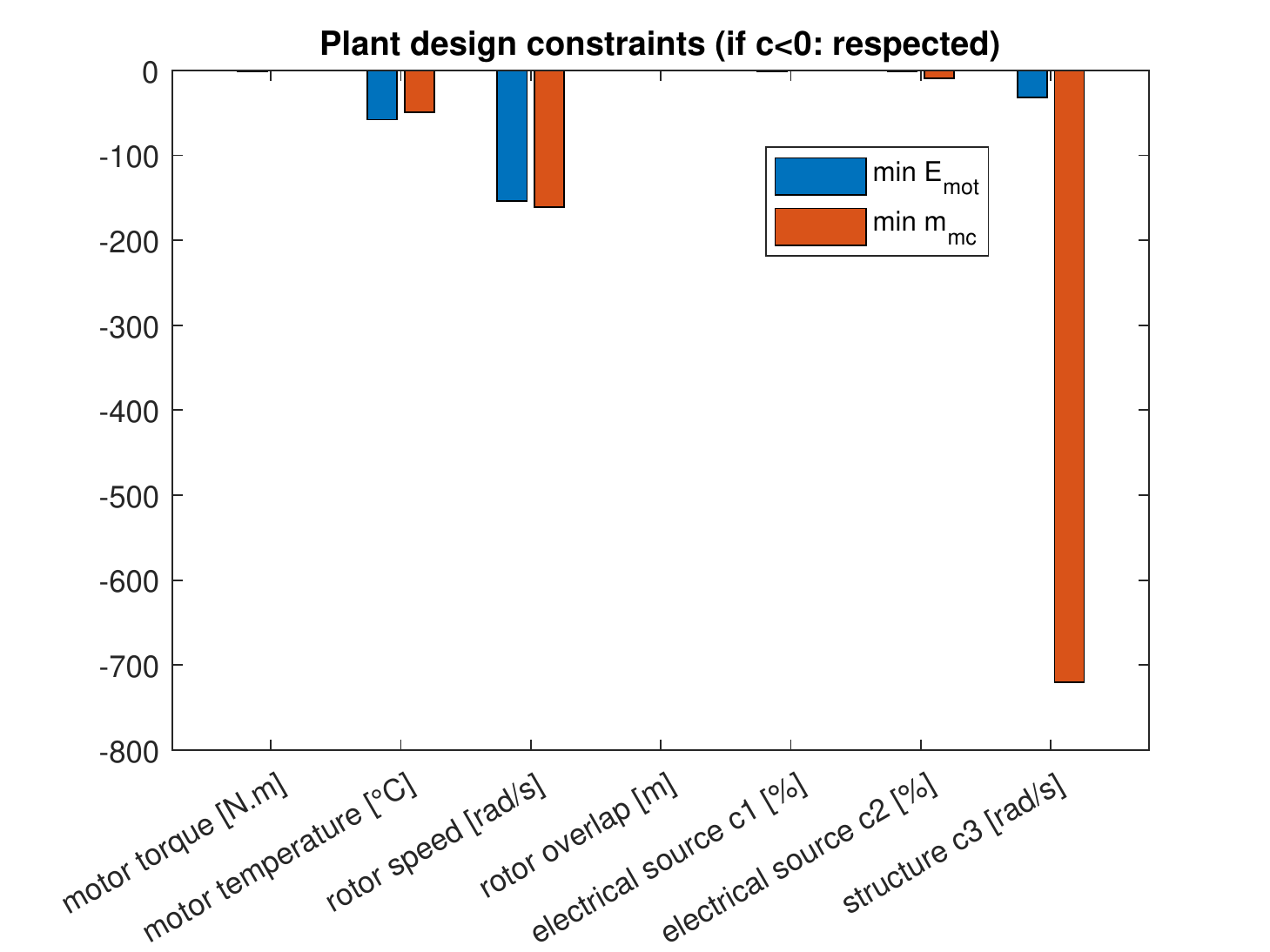}
    \caption{Design constraints after optimization}
    \label{fig:const_optim}
\end{figure}

The design constraints are represented on the figure~\ref{fig:const_optim}. By convention, a negative or zero value means that the constraint is satisfied. Hopefully, after the optimization, all constraints are satisfied. The constraints at the border/edge of the feasible domain are 
rotor overlap, motor torque and electrical source constraints.
These constraints are limiting constraints. In other words, only the constraints on 
rotor speed, motor temperature and structural constraint 
are not reached.

\begin{figure}[htp]
    \begin{subfigure}{\textwidth}
    \centering
    \includegraphics[width=0.55\textwidth]{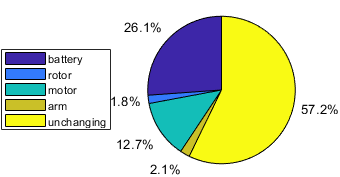}
    \caption{Reference model}
    \end{subfigure}
    \hfill
    \begin{subfigure}{0.49\textwidth}
    \centering
    \includegraphics[width=0.85\textwidth]{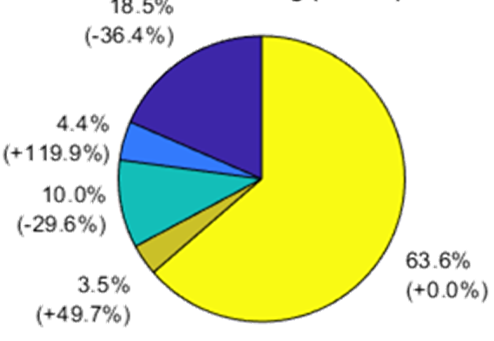}
    \caption{Optimized model for energy reduction}
    \end{subfigure}
    \hfill
    \begin{subfigure}{0.49\textwidth}
    \centering
    \includegraphics[width=0.85\textwidth]{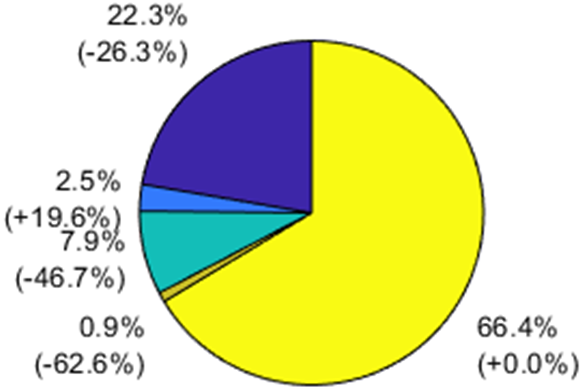}
    \caption{Optimized model for mass reduction}
    \end{subfigure}
    \caption{Mass distribution and (relative change between parentheses)}
    \label{fig:mass_repart}
\end{figure}

The distribution of the mass into five parts is shown in figure~\ref{fig:mass_repart}. The most important part is the unchanging mass of components independent of the design variables. It includes for instance the frame, the payload...and represents 57\% of the total mass in the reference model and the relative change (number between parentheses) is null for the two models after energy and mass optimizations. The battery mass is more reduced for the design where energy is minimized (36\%) than for the one where mass is minimized (26\%), as shown previously. When the energy consumed is minimized, the rotors and arms masses are significantly increased, as opposed to the motor mass which decreases to 30\%.

\begin{figure}[htp]
    \centering
    \includegraphics[width=0.7\textwidth]{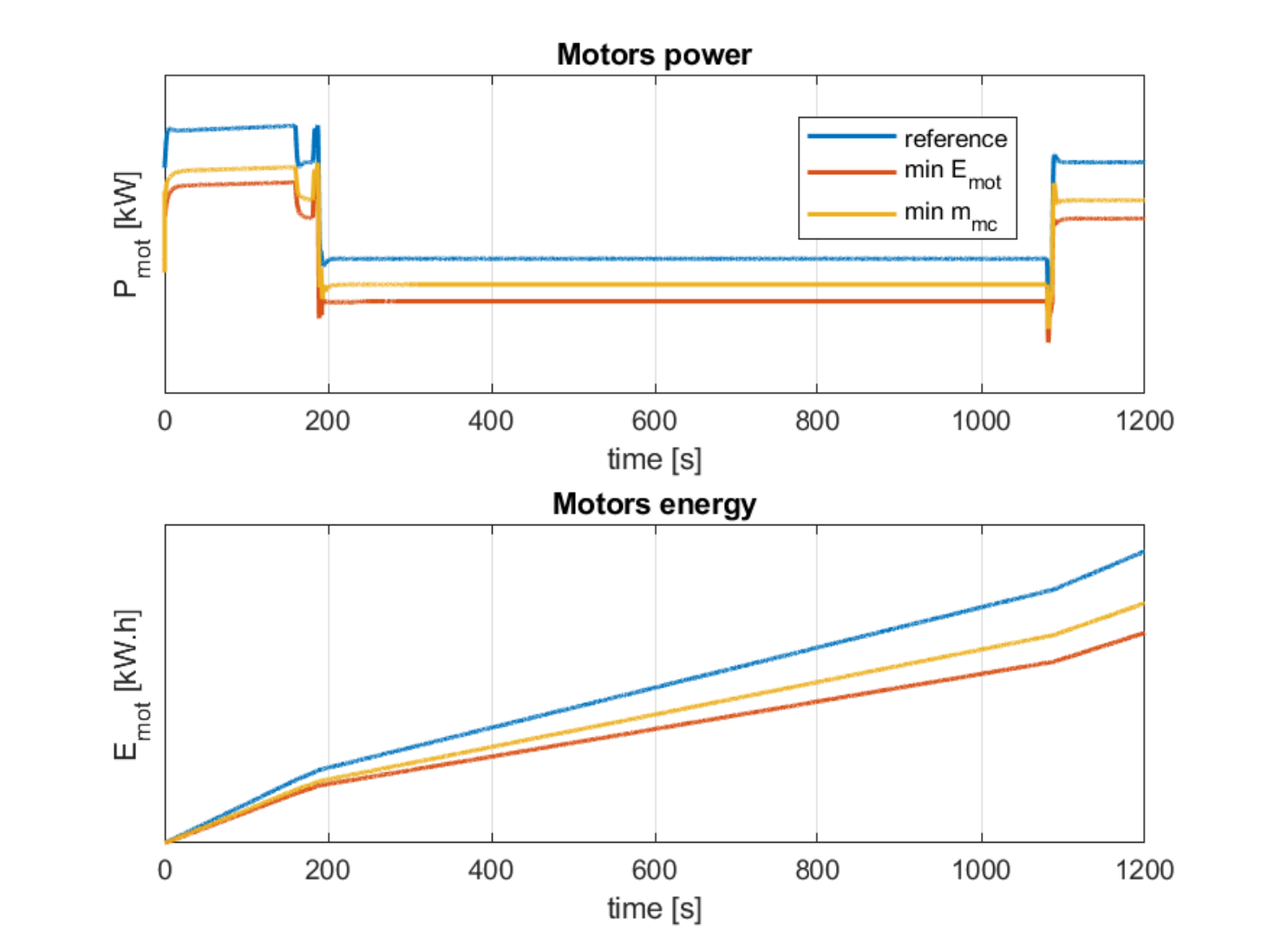}
    \caption{Motor power and energy during a standard mission for three different models\\ \footnotesize For confidentiality reasons the value of power and energy are not shown}
    \label{fig:motor_power_energy}
\end{figure}

The motor power and energy during a standard mission for the three different models (reference, optimized for energy reduction and optimized for mass reduction) are shown in figure~\ref{fig:motor_power_energy}. At the end of the mission, as expected, the lowest motor energy is that of the model designed for minimizing the energy consumed for this mission, the optimized multicopter allows an energy gain of 37\%. This configuration also corresponds to the lowest motor power.

\subsection{Passive Fault tolerant Control}

\begin{figure}[htp]
    \centering
    \includegraphics[width=0.65\columnwidth]{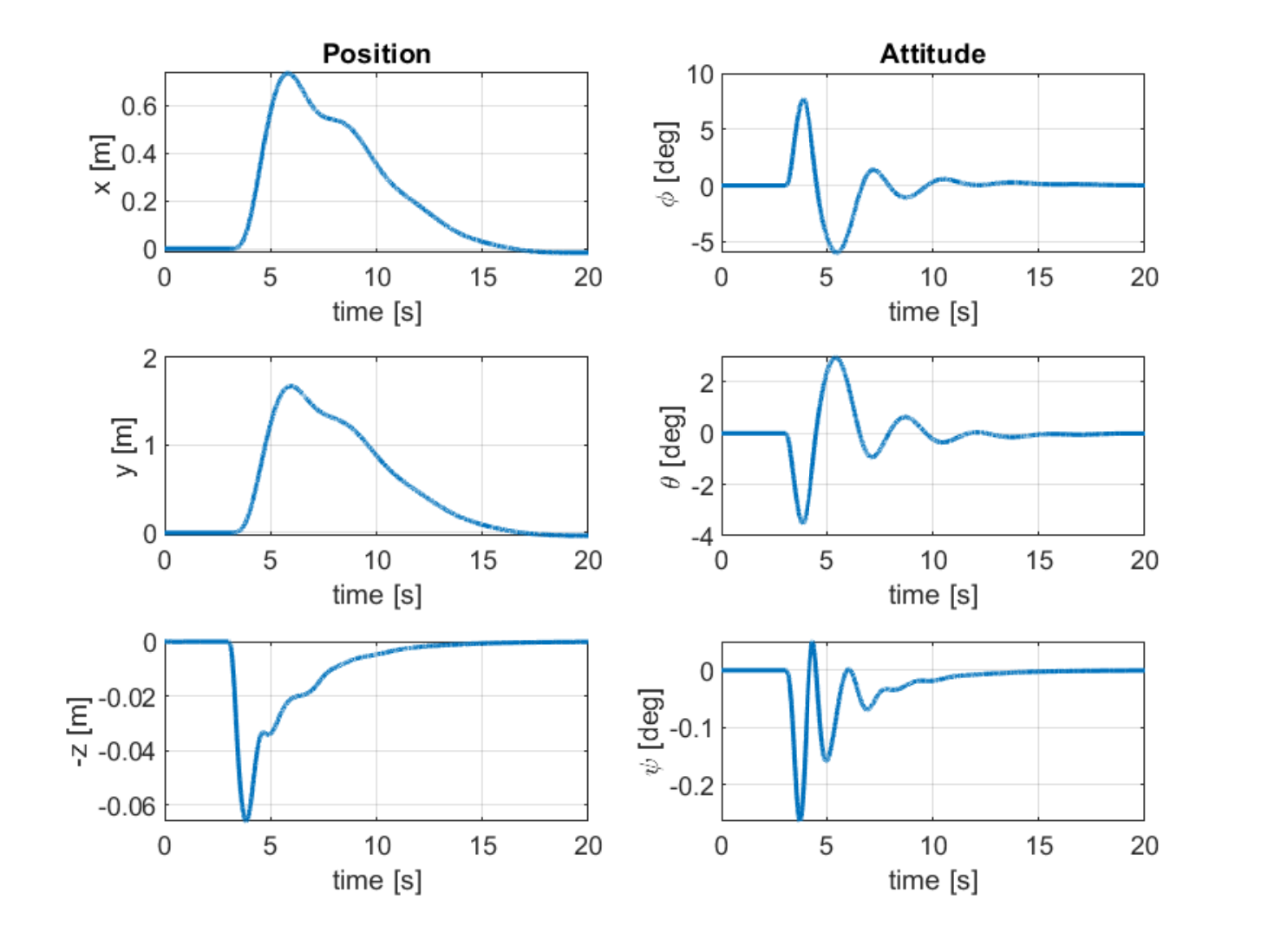}
    \caption{Position and attitude: structured $H_\infty$ synthesis}
    \label{fig_xyz_angle_passive_FTCS_Hinf}
\end{figure}

\begin{figure}[htp]
    \centering
    \includegraphics[width=0.75\columnwidth]{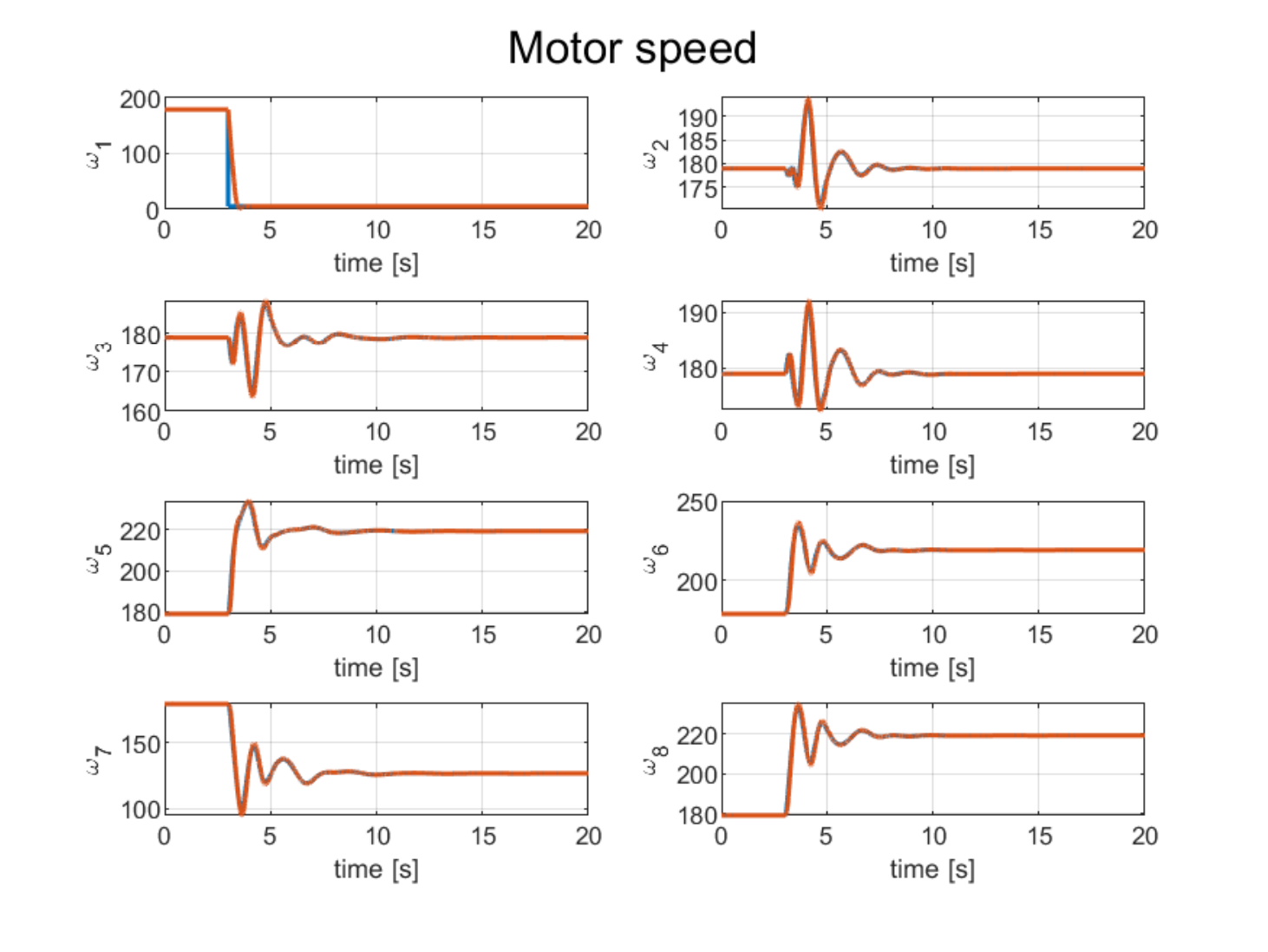}
    \caption{Motor speed: structured $H_\infty$ synthesis, rotor 1 fails}
    \label{fig_mot_speed_passive_FTCS_Hinf}
\end{figure}

Following the strategy of passive Fault tolerant Control introduced in Section 3, the co-design optimization is carried out considering models with and without failure in order to meet the robustness requirements of the control laws.
The passive FTC is validated on the optimized vehicle in a faulty situation. The multicopter is in hovering flight when the first rotor fails at three second of the simulation. Fig.~\ref{fig_xyz_angle_passive_FTCS_Hinf} and Fig.~\ref{fig_mot_speed_passive_FTCS_Hinf} represent respectively the position and attitude of the multicopter and the rotational speed of each rotor. This robust controller ensures stable behavior for any single rotor failure. Indeed, the multicopter reaches a stabilized hovering flight within 10 seconds. In Fig.~\ref{fig_mot_speed_passive_FTCS_Hinf}, the rotational speed of rotor 1 falls to zero after the fault. To compensate the lack of thrust of the first rotor, its opposite rotor (the $7^{th}$) is slowed down. The other rotors give a higher or at least the same thrust in steady state after the fault event in order to offset the gravity.

\newpage
\section{Conclusions}
\label{sec:concl}

In this paper, a co-design approach was proposed and applied to size an energy-efficient multicopter robust to failure. The proposed method is based on a bi-level optimization with two loops, one for the multicopter design parameters and one for the control gains tuned for the vehicle with or without failures. The bi-level strategy was necessary because the choice of robust $H_\infty$ control optimization alone was not compatible with the integration of nonlinear temporal response constraints. This approach relies on the ability to develop and assemble sizing models of the main multicopter components. In this paper, sizing models are established using scaling laws, which is a very powerful tool for sizing components from reference components during preliminary design phases. The first results have shown a significant reduction in the consumed energy, which is the main objective, for the multicopter in all conditions. Further work will be achieved to analyze the sizing for different flight phases and to include the effect of a failure case in the sizing loop in addition to the failure cases already integrated in the control loop. Also, the approach will be adapted to include additional dynamic performance scenarios such as handling qualities. In addition, the capability to use parallelization on certain tasks of the co-design process will be investigated to reduce further the computation times.

\section{Contact Author Email Address}
mailto: jordan.mabboux@isae-supaero.fr

\section{Copyright Statement}
\begin{small}
The authors confirm that they, and/or their company or organization, hold copyright on all of the original material included in this paper. The authors also confirm that they have obtained permission, from the copyright holder of any third party material included in this paper, to publish it as part of their paper. The authors confirm that they give permission, or have obtained permission from the copyright holder of this paper, for the publication and distribution of this paper as part of the ICAS proceedings or as individual off-prints from the proceedings.
\end{small}

\newpage

\biblio{references}

\end{document}